\documentstyle[prl,aps,multicol,epsf]{revtex}
\begin{document}
\title{Jahn-Teller distortion and electronic correlation effects in undoped 
manganese perovskites}
\author{Patrizia Benedetti and Roland Zeyher}
\address{Max-Planck-Institut\  f\"ur\
Festk\"orperforschung, Heisenbergstrasse 1, 70569 Stuttgart, Germany \\}

\date{\today}
\maketitle
\begin{abstract}
The formation of a long-range Jahn-Teller distortion in undoped manganites
is studied within the dynamical mean-field theory taking both
electron-phonon and electron-electron interactions into account. We find
that the observed insulating behavior at temperatures above the Jahn-Teller transition 
and the bond-length changes below it can most naturally
by explained by assuming a strong electronic repulsion and a rather weak
electron-phonon coupling. 
\end{abstract}
\begin{multicols}{2}

\tightenlines
The physics of manganese oxides Re$_{1-x}$A$_x$MnO$_3$ (Re is a rare earth
metal ion such as La or Pr, A a divalent alkali such as Sr or Ca)
is characterized by strong electronic \cite{Varma} and electron-phonon 
\cite{Solovyev,Millis1} interactions.
Due to the mainly local nature of these interactions the Dynamical 
Mean-Field (DMF)  theory \cite{Kotliar,Metzner} seems to be an 
appropriate tool 
for investigations of these systems. In the following we will extend 
previous treatments \cite{Millis1,Millis2,Nagaosa} by including both 
electron-phonon and electron-electron interactions and
also by allowing for phases with long-range order. We will confine 
ourselves to the undoped case $x=0$ and study, in particular, the 
question whether the observed Jahn-Teller distortion is driven more by the 
electron-phonon coupling or by electronic correlations.

Taking the limit of a very large Hund's rule coupling between the core
3/2 spins and the $e_g$ electrons the resulting simplified 
Hamiltonian reads \cite{Maezono} 

\begin {eqnarray}
H &=& -\sum_{ij\alpha \alpha '}t_{ij}^{\alpha \alpha '} 
(c^{+}_{i \alpha}c_{j\alpha'}+H.c.)
-  \frac{U}{2} \sum_{i,\gamma}T_{i \gamma}T_{i \gamma}\nonumber \\
&+& \frac{1}{2} \sum_{i \gamma}[ (\partial_{t}\phi_{i \gamma})^{2}+
\omega_{0}^{2}\phi_{i \gamma}^{2}]
+g \sum_{i \gamma} \phi_{i \gamma}T_{i \gamma},
\end{eqnarray}
with
\begin {eqnarray}
T_{i \gamma}=\sum_{\alpha \alpha'} c^{+}_{i \alpha} \sigma_{\gamma}
^{\alpha \alpha'}c_{i\alpha'}.
\end{eqnarray}
$c^{+}_{i \alpha} (c_{i \alpha})$ creates (destructs)
an electron on the $i$th Mn site in the $e_{g}$ orbital $\alpha=\uparrow 
\downarrow$, with $|\uparrow \rangle=d_{x^{2}-y^{2}}, |\downarrow \rangle=d_{3z^{2}-r^{2}}$. 
Because the Hund's rule coupling constant has assumed to be infinitely
large there are no spin indices in $H$. The hopping elements $t$
depend in the DMF theory on the background of the core spins. 
In the case of a ferromagnetic background $t$ is given by the bare hopping
element, for a paramagnetic background $t$ contains reduction factors
\cite{Millis1} due to the double exchange mechanism. In the following
we take for simplicity 
$t_{ij}^{\alpha \alpha '}=\delta_{\alpha \alpha '}t_{ij}$. 
The pseudospin operator $T$ is defined in Eq. (2) where the index $\gamma$
always runs only over $\gamma = x,z$ where $x$ and $z$  denote the
corresponding Pauli matrices in Eq. (2). The effective Hubbard constant
$U$ is equal to $U'-J$, where $U'$ is the inter-orbital Coulomb and
$J$ the exchange constant of the two $e_g$ orbitals. $\phi_{i \gamma}$
denote the two normal Jahn-Teller modes of the six oxygens surrounding
the $i$th Mn site and having the local $e_g$ symmetry. $\partial_t$ is
the time derivative and $\omega_0$ and $g$ are the frequency and the
coupling constant of the phonons. 

The Hamiltonian Eq. (1) together with the condition $t_{ij}^{\alpha \alpha '}=\delta_{\alpha \alpha '}t_{ij}$, is not only invariant under the
transformations of the octahedral point group but also under infinitesimal rotations
in orbital space. This larger symmetry is due to the fact that $H$
contains only invariants which are bilinear in the fields. 
Using the rotational 
symmetry we can thus omit components with  $\gamma=x$ in Eq. (1) within
a good approximation, yielding \cite{Nagaosa}
\begin {eqnarray}
H &=& -\sum_{ij\alpha }t_{ij} (c^{+}_{i \alpha}c_{j\alpha}+H.c.)
+ U \sum_{i}n_{i \uparrow}n_{i\downarrow}
\nonumber \\
&+& \frac{1}{2} \sum_{i}[ (\partial_{t}\phi_{i})^{2}+
\omega_{0}^{2}\phi_{i}^{2}]
+g \sum_{i} ( n_{i \uparrow}-n_{i \downarrow})\phi_{i},
\end{eqnarray}
where $\phi_i$ denotes the field $\phi_{iz}$.

We want to investigate also the case where the orbitals order in a 
long-range anti-ferro orbital pattern. We therefore allow for broken symmetry 
and divide the lattice into two inequivalent sublattices
A and B. Within the DMF theory the action associated with the Hamiltonian (1)
reduces to two single impurity actions $S_{eff,i}$ where $i=A,B$ and $A$ and
$B$ stand for one site on the sublattices $A$ and $B$, respectively.
The two impurity actions read
\begin {eqnarray}
S_{eff,i} &=&\int^{\beta}_{0}d\tau \int^{\beta}_{0}d\tau'\sum_{\alpha=\uparrow
\downarrow}c_{i \alpha}^{*}(\tau)G_{0 i \alpha}^{-1}(\tau-\tau')c_{i \alpha}(\tau')
\nonumber \\
&-&\frac{1}{2}
\int^{\beta}_{0}d\tau\,\phi_i(\tau)\Big(-\frac{d^{2}}{d\tau^{2}}
+\omega_{0}^{2}\Big)\phi_i(\tau)\nonumber \\
&-&g\int^{\beta}_{0}d\tau\, [n_{i \uparrow}(\tau)-n_{i \downarrow}(\tau) ]
\phi_i(\tau)  \nonumber \\
&-&U\int^{\beta}_{0}d\tau\, n_{i \uparrow}(\tau)n_{i \downarrow}(\tau).  
\end{eqnarray}

The two local Green's functions are related by
\begin{equation}
G_{A \alpha}=G_{B \bar{\alpha}},
\end{equation}
where $\bar{\alpha}$ is the orbital different from $\alpha$.
The relations between the Weiss field $G_{0i\alpha}$ and the local Green's 
function become in the case of the Bethe lattice \cite{Kotliar}   
\begin{eqnarray}
G_{0 A {\alpha}}^{-1}&=&i\omega_{\nu}+\mu-t^{2}G_{B {\alpha}}\nonumber\\
G_{0 B {\alpha}}^{-1}&=&i\omega_{\nu}+\mu-t^{2}G_{A {\alpha}}.
\end{eqnarray}
Decoupling the last term in Eq. (4) by means of a Hubbard-Stratonovich
transformation and integrating out the electronic degrees of freedom
we obtain the new impurity actions 

\begin {eqnarray}
S_{eff,i} &=& -\frac{1}{2} \int^{\beta}_{0}d\tau\,\phi_i(\tau)
\Big(-\frac{d^{2}}{d\tau^{2}}+\omega_{0}^{2}\Big)\phi_i(\tau)
\nonumber \\
&-&\frac{1}{2} \int^{\beta}_{0}d\tau\;y_i^{2}(\tau) 
+ Tr\,\ln[G_{0 i \uparrow}^{-1}-(g \phi_i+\sqrt{U}y_i)]
\nonumber \\ 
&+&Tr\,\ln[G_{0 i \downarrow}^{-1}+(g \phi_i+\sqrt{U}y_i)].
\end{eqnarray}
$y_i(\tau)$ denotes the Hubbard-Stratonovich (HS) field. Using Eq. (7)
the local Green's functions have the exact representations

\begin{eqnarray}
G_{i\alpha}(\tau &-& \tau') = \int\,D\phi_i(\tau) Dy_i(\tau) 
\,[G_{0i\alpha}^{-1}(\tau_{1}-\tau_{2})- \nonumber \\
\sigma_{z}^{\alpha \alpha}(g\phi_i(\tau_{1})  
&+&\sqrt{U}y_i(\tau_{1}))\delta(\tau_{1}-\tau_{2})]^{-1}_{\tau-\tau'} 
e^{S_{eff,i}}/Z_i, 
\end{eqnarray}
for $i=A,B$. $Z_i$ is the partition function.
Eqs. (5)-(8) represent a closed system of equations for the Green's 
functions of the two coupled sublattices.

From the  equation of motion for the field $\phi_i$
one obtains
\begin{eqnarray}
\langle \phi_i \rangle &=&-\frac{g}{\omega_{0}^{2}} 
(\langle n_{i \uparrow} \rangle - \langle n_{i \downarrow}\rangle),  
\nonumber \\
\langle \phi_A \rangle_ &=&-\langle \phi_B \rangle. 
\end{eqnarray}
Here $\langle n_{i \alpha}\rangle$ is the average number of electrons in 
the orbital $\alpha$ on the sublattices $i=A,B$. The quantitiy 
$\langle n_{i \uparrow} \rangle - \langle n_{i \downarrow}\rangle$
can be considered as a staggered order parameter.
 $\langle \phi_i \rangle$ is the 
expectation value of the phonon field $\phi_i$. Moreover, 
the average number of electrons per manganese site is one in the undoped
case, so that $n_i=\langle n_{i \uparrow}+n_{i \downarrow}\rangle=1$. \par
Assuming that the two fields $\phi$ and $y$ vary only slowly in time 
$S_{eff,i}$ can be calculated by means of a gradient expansion \cite{cond-mat}.
The leading contribution to $S_{eff,i}$ is local in time and is given
by the following effective potential
\begin{eqnarray}
&\,&V_i(\phi(\tau),y(\tau))= \frac{\omega_{0}^{2}}{2} \phi_i^{2}(\tau)
+\frac{1}{2}y_i^{2}(\tau) \nonumber \\
&-&T \sum_\nu e^{i \nu0{^+}} \ln[1-(g\phi_i(\tau)+\sqrt{U}y_i(\tau)) 
G_{0 i\uparrow}(i\nu)] \nonumber \\
&-&T \sum_\nu e^{i \nu0{^+}} \ln[1+(g\phi_i(\tau)+\sqrt{U}y_i(\tau)) 
G_{0 i\downarrow}(i\nu)]. 
\end{eqnarray}
The lowest-order non-local contributions to $S_{eff,i}$ are functionals of
$\phi_i$ and $y_i$ which include also at least one time derivative ${\partial
\phi_i}/{\partial \tau}$ or ${\partial y_i} /{\partial \tau}$.

For the evaluation of $G$ in Eq. (8) we distinguish between several cases.
First we consider the case $U=0$ where the model reduces to a pure
Jahn-Teller system and the collective coordinate $y$ drops out. 
Starting from an educated guess for $G_{0i\alpha}$ we calculate first
the effective potentials $V_i$ using Eq. (10) and transforming the sums
over Matsubara frequencies into integrals along the real axis. $G_{i\alpha}$,
as given by Eq. (8), is calculated by using expansions around the extremal
paths of $S_{eff,i}$. In general, there are two time-independent paths
$\bar{\phi}_r,r=1,2$, corresponding to the two minima of $V_i$,
and one time-dependent instanton path. The contribution from the latter is
very small for our considered temperatures and atomic masses \cite{cond-mat} 
so we will neglect it together with the non-local part of $S_{eff,i}$. 
Around each minimum $\bar{\phi}_r$ we generate a
Migdal-Eliashberg-like expansion. It is governed by the  Migdal
parameter $\omega_0/t $ which is rather small in the manganites. 
Detailed calculations show \cite{cond-mat} that 
for all  finite coupling strengths it is sufficient to include only 
the Hartree and the Fock terms. First-order vertex corrections to 
these terms turn out to be smaller by a factor $\lambda \omega_0/t$, 
$\lambda$ being the electron-phonon coupling constant
defined by $\lambda=g^2/\omega_0^2t$. In calculating the Fock term we 
use a phonon propagator
which has been constructed from the exact eigenvalues and eigenfunctions
of the Schr\"odinger equation describing an atom in the effective potential
$V_i$. In this way anharmonic effects are fully taken into account.
From Eq. (6) we obtain new Weiss fields $G_{0i\alpha}$ as input for 
another cycle. The procedure is repeated until self-consistency is reached.
 From now on all energies are expressed in units of the hopping matrix 
element $t$, unless differently specified.
\vspace{-0.5cm}
\begin{figure}
      \epsfysize=75mm
      \centerline{\epsffile{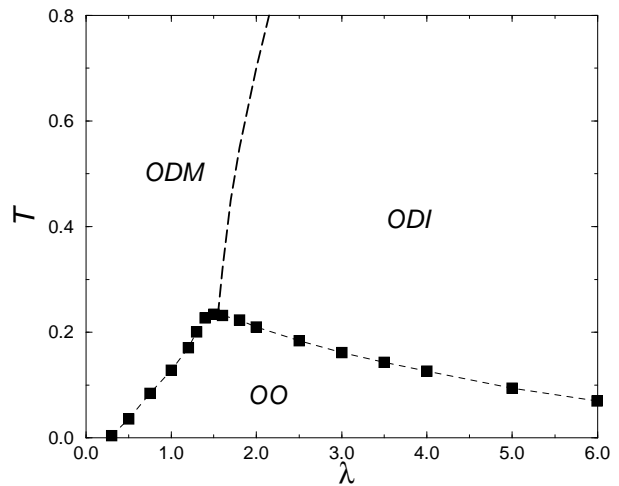}}
\narrowtext
\caption{
T-$\lambda$ phase diagram for electrons coupled only to Jahn-Teller 
phonons. OO: ordered orbital phase; ODM and ODI: metallic and insulating
disordered orbital phases, respectively.}
\label{fig:phase1}
\end{figure}

Numerical results are shown in Fig. \ref{fig:phase1} where 
the critical temperatures are plotted  versus the electron-phonon coupling $\lambda$. 
The phonon frequency has been fixed 
to $\omega_{0}=0.04$. Three phases can be identified:
1) an insulating phase in which the orbitals are ordered (OO), 2)  a phase in which 
the orbitals are disordered and the system is metallic (ODM), 3) a phase in 
which the orbitals are disordered and the system is insulating (ODI).
The thick dashed line shows the metal-insulator transition temperature, 
the squares indicate the calculated points where the orbital order-disorder 
transition takes place. This
is a second-order phase transition where the order parameters 
$\langle \phi_i \rangle$ go continuously to zero as the temperature 
approaches $T_{c}$ from below.

In the orbital disordered phase the two 
$e_{g}$ orbitals are degenerate and occupied with the same probability.
The effective potentials $V_i$ are symmetric with respect to $\phi_i=0$
and two regimes can be distinguished. 
In the strong coupling regime,  $\lambda > \lambda_{c}(T)$, the potential 
shows two deep and well separated minima
leading to a gap in the density of electronic states.
In this case the system is in the insulating phase ODI. In the weak 
coupling regime, $\lambda < \lambda_{c}(T) $, 
the potential has only one minimum and the system is in the metallic 
phase ODM. 
In the ordered phase the effective
potential of each sublattice is asymmetric with respect to $\phi_i=0$.
In the weak coupling regime there is  only one minimum and it
is shifted with respect to the origin by an amount 
$\Delta \phi_i \sim  \langle \phi_i \rangle$. In contrast
to that  in the strong coupling case such a shift is also accompanied by 
a lowering of one minimum with respect to the other minimum. 
The interaction with the phonons lifts the orbital degeneracy in an 
opposite way on the two sublattices.
This creates a  disproportion in the occupation probability of the two orbitals
leading to the formation of a staggered pattern of occupied orbitals and consequently to the experimentally observed
cooperative distortion of the oxygen bonds. A qualitatively similar phase
diagram has recently been given in Ref.\cite{ciuchi} for the Holstein model  
with infinitely heavy atoms.
 
Experimentally the static Jahn-Teller distortion in PrMnO$_3$ occurs around
$T_c \sim 850K$ \cite{Jirak} and in LaMnO$_3$ around $T_c \sim 750K$.
Above $T_c$ these systems are insulators.
If we compare these data with our phase diagram in Fig. \ref{fig:phase1} 
we obtain, taking $t=0.7 eV$ \cite{Saitoh},
a value for $\lambda$ of about $5$ in the insulating region. This is an
extremely large value.
In the following we show that a more realistic picture is obtained if one 
also includes the Coulomb repulsion between the $e_{g}$ electrons.

$S_{eff,i}$ depends on the two fields $\phi_i$ and $y_i$ if also the 
electron-electron 
interaction is included. A reduction to one field is in this case
possible if one treats the phonon field in the static approximation.
The physical basis for this is the inequality $\omega_{0} \ll t$, 
which says that the effective phonon mass is much larger than the
electronic one. Performing then the following change of variables in 
Matsubara space 
\begin {eqnarray}
y_{i}(i\omega_n)\sqrt{U}+g\phi_{i}(0)\,\delta_{n,0}=x_i(i\omega_n)
\sqrt{U+\lambda},
\end{eqnarray}
we can integrate out the $\phi_i(0)$ coordinate and obtain the effective 
action
\begin {eqnarray}
S_{eff,i} &=& -\frac{1}{2}\sum_{n}
\left(1+\frac{\lambda}{U}-\frac{\lambda}{U}\delta_{n,0}\right)x_i(i\omega_n)
x_i(-i\omega_n) 
\nonumber \\ 
&+& Tr\,\ln[G_{0i\uparrow}^{-1}-\sqrt{U+\lambda}\,x_i] \nonumber \\
&+&Tr\,\ln[G_{0i\downarrow}
^{-1}+\sqrt{U+\lambda}\,x_i].
\end{eqnarray}
The limit $\lambda \rightarrow 0$ correctly reproduces the action for the 
Hubbard model
and the limit $U \rightarrow 0$ the action for the static Jahn-Teller 
system. In the following we will concentrate on the case $U \gg \lambda$. 
Using the local approximation for the $Tr\,\ln$ term the effective 
potential reads
\begin {eqnarray}
&\,&V_i(x_i(\tau))= \frac{1}{2} x_i(\tau)^{2}\nonumber \\ 
&-&T \sum_\nu e^{i \nu0{^+}} \ln[1-\sqrt{U+\lambda}\,x_i(\tau)G_{0i\uparrow}(i\nu)]\nonumber \\ 
 &-&T \sum_\nu e^{i \nu0{^+}} \ln[1+\sqrt{U+\lambda}\,x_i(\tau) G_{0i\downarrow}
(i\nu)]. 
\end{eqnarray}
This means that there is no kinetic term in Eq. (12). $G_{i\alpha}$ can 
again be calculated by using expansions around the potential
minima $\bar{\phi}_r$. The propagator for the field $x$ becomes instantaneous
and thus can be put to zero. As a result only the Hartree term in each
expansion around $\bar{\phi}_r$ contributes. The lowest-order corrections
to the field propagator would come from the non-local part in 
$S_{eff,i}$ and lead to a relaxational dynamics for the field $x$.
Since for all couplings the damping is substantially smaller than 
$\omega_0$ these corrections are, however, small \cite{cond-mat}. 
\vspace{-0.5cm}
\begin{figure}
      \epsfysize=75mm
      \centerline{\epsffile{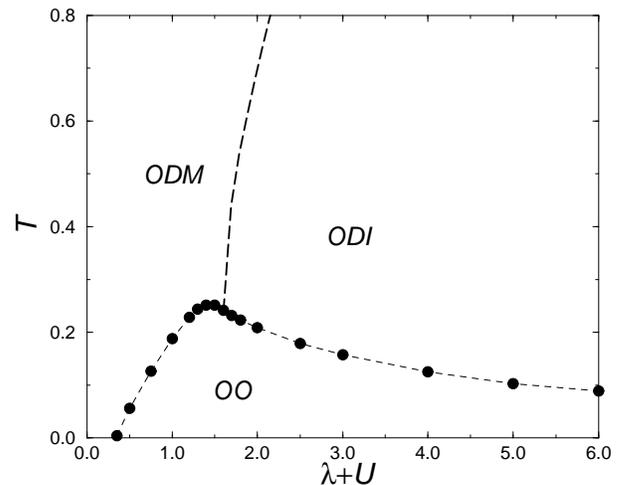}}
\narrowtext      
\caption{
Phase diagram showing the critical temperature  versus 
$\lambda + U$ for a static phonon and dynamic HS field assuming 
$U \gg \lambda$.} 
\label{fig:phase_a}
\end{figure}

 We have solved Eqs.(11)-(13) in a self-consistent way. 
Figure \ref{fig:phase_a} shows the calculated phase 
diagram for the case where
the phonon, but not the Hubbard-Stratonovich field, is treated in the 
static approximation. We also assumed $U \gg \lambda$ which allows
to lump the two coupling constants together into an effective
coupling constant $\lambda + U$. The black 
dots are the calculated critical temperatures as a function of
$\lambda+U$. 
Our previous discussion of the three phases OO, ODM, and ODI in Fig. \ref{fig:phase1}
also applies more or less to Fig. \ref{fig:phase_a}. Remarkable is
the close similarity of the phase boundaries in Figs. \ref{fig:phase1}
and \ref{fig:phase_a}, even on a quantitative level.
Let us estimate the Jahn-Teller coupling in the present case.
Using $t=0.7eV$, $T_c\sim 0.1t$ and the fact that LaMnO$_3$ is
insulating above $T_c$ we obtain from Fig. \ref{fig:phase_a} 
an effective coupling $\lambda+U \sim 5.3$ and 
$\Delta n=n_{A \uparrow}-n_{A \downarrow}\sim 1$.
The energy associated with the Jahn-Teller distortion can be estimated 
from the experimental data  
and is about $E \sim 0.14 eV$. From that, making use of Eq. (9) with 
$\Delta n \sim 1$, we obtain $\lambda \sim 0.3$.
This means that the experimental data can easily be reproduced with
$U\sim  3.5 eV$ and $\lambda \sim 0.3$. An independent determination of $U$
yielded $3.8 eV$ \cite{Feiner}  and  is thus
in good agreement with our value. The value $\lambda = 0.3$
associated with one branch of phonons is on the one hand side not 
small indicating the importance of phonons. On the other hand, the
critical coupling strength for a metal-insulator transition is according to
Fig. \ref{fig:phase1} about 1.5 and thus five times larger than the
obtained value. This means that the insulating phase above $T_c$ 
as well as the Jahn-Teller transition itself in
PrMnO$_3$ and LaMnO$_3$ are mainly caused by electronic correlations
and not by the electron-phonon coupling.

\vspace{-0.5cm}
\begin{figure}
     \epsfysize=80mm
      \centerline{\epsffile{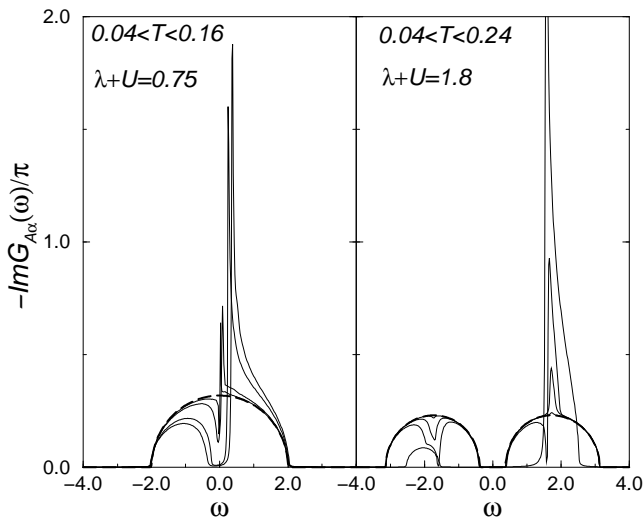}}
\narrowtext     
\caption{ Temperature dependence of the electronic density of states 
on the sublattice A for the orbital $\alpha = \uparrow$ versus frequency. 
The left and right panels
correspond to the weak- and strong-coupling cases, respectively.}
\label{fig:Ga}
\end{figure}

Fig. \ref{fig:Ga} shows the evolution of the electronic density of states (DOS)
of one $e_{g}$ orbital on the sublattice A with temperature for two 
different values of the 
effective coupling constant $\lambda +U$. The solid lines cover the 
temperature range from far below the orbital
ordering temperature $T_{c}$ to just
below $T_{c}$. The dashed line shows the DOS just above $T_{c}$.
The left panel presents data in the weak-coupling regime $\lambda +U=0.75$. 
Below the critical temperature the DOS shows an 
asymmetry with respect to the Fermi level $\omega=0$, signalizing a 
disproportion of the average 
electronic occupation of each $e_{g}$ orbital. As a result, 
$n_{A \uparrow} \neq n_{A \downarrow}$ and, at the same time,  
$n_{A \uparrow} = n_{B \downarrow}$ holds. 
The difference between $n_{A \uparrow}$ and  $n_{A \downarrow}$
increases with decreasing temperature. At low enough 
temperature
a gap opens in the DOS at $\omega=Re\,G_{B \alpha}(\omega)
+\sqrt{\lambda +U}\langle x_A \rangle$.
$Re\,G_{B \alpha}(\omega)$ denotes the real part of $G_{B \alpha}(\omega)$ 
and $\langle x_A \rangle$
the expectation value of the field $x$ for the action Eq. (12) on the 
sublattice A. 
Above $T_{c}$ the system is metallic.\\
The right panel in Fig. \ref{fig:Ga} shows 
data for  $\lambda +U=1.8$. Above $T_c$ the system 
is in the strong-coupling regime where the effective potential has two well 
separated minima causing two symmetric subbands in the DOS.
With decreasing temperature spectral weight is transferred from the lower to 
the upper band leading to an asymmetry in the spectra. 
The gap stays practically constant until, at low enough temperatures,
the orbital ordering gap becomes larger 
than the charge gap and almost all spectral weight is transferred to the 
right band. At T=0.04t we have $\Delta n=n_{A \uparrow}
-n_{A \downarrow}\sim 0.86$.\par
\begin{figure}
      \epsfxsize=120mm     
      \centerline{\epsffile{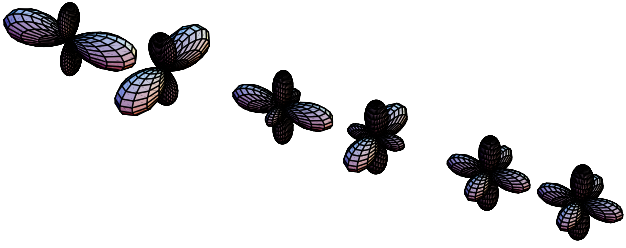}}
\vspace{-8.0cm}
\narrowtext
\caption{Angular dependence of the electronic density
on two adjacent Mn sites for three different temperatures
and $\lambda + U = 4.0$.  The vertical lobe points into the $c$-direction,
the horizontal ones alternatively into the $a$- and $b$-directions.}
\label{fig:orb}
\end{figure}

 
As mentioned previously our Hamiltonian Eq. (1) is for the case of isotropic
hopping matrix elements invariant under continuous rotations in orbital
space which can be characterized by an angle $\theta$\cite{Maezono}.
As a result $\theta$ is not determined by Eq. (1) but by residual interactions
such as anisotropic hoppings, higher-order invariants in the couplings
due, for instance, to a quadratic electron-phonon coupling. The real
situation in the manganites is even more complicated because of the 
observed tilt of the oxygen octahedra and the important role  played 
by the oxygen degrees of freedom. However, we can discuss with our 
Hamiltonian the temperature 
dependence of the electron density on the Mn sites if we choose, 
for instance,
$\theta=\pi/4$ which is not far away from the experimental situation
in PrMnO$_3$ and LaMnO$_3$. 
Fig. \ref{fig:orb} shows for $\lambda + U = 4.0$ the angular dependence of 
the electronic density on the two sublattices i=A,B, defined by
\begin{equation}
\rho_i(\theta,\varphi)=n_{i\uparrow}|\langle\uparrow|\theta,\varphi \rangle|^{2}
+n_{i\downarrow}|\langle\downarrow|\theta,\varphi \rangle|^{2},
\end{equation}
$\langle\uparrow|\theta,\varphi \rangle$ and  $\langle\downarrow|\theta,\varphi \rangle$ denote 
the angular components of 
$|\uparrow \rangle=(d_{x^{2}-y^{2}}+d_{3z^{2}-r^{2}})/\sqrt{2}$ and  
$|\downarrow \rangle=(d_{x^{2}-y^{2}}-d_{3z^{2}-r^{2}})/\sqrt{2}$, respectively.

\vspace{-2.0cm}

The first pair of densities on the left in Fig. \ref{fig:orb} corresponds
to the temperature $T=0.04t< T_{c}$ and the site
occupancies $n_{A\uparrow}=0.98$ and $n_{B\uparrow}=1-n_{A\uparrow}=0.02$. 
The electronic charge
distribution is strongly anisotropic leading to alternating short and long
Mn-O bonds along the $a$- and $b$- directions on the two sublattices.
The bond in the $c$-direction remains constant. The pair of densities in
the middle of Fig. \ref{fig:orb} corresponds to  $T=0.116t < T_{c}$ 
and the site occupancy $n_{A\uparrow}=0.67$. One recognizes that
the increasing temperature decreases the anisotropy
of the electronic distribution and thus also the distortion of the octahedra.
Finally, the pair of densities at the right depicts the electron distribution
on the two sublattices for $T=0.16t > T_{c}$
and $n_{A\uparrow}=n_{B\uparrow}=0.5$. The electrons are now uniformally 
distributed along the three axes and the oxygens occupy the equilibrium 
positions of the cubic perovskite lattice. 

The above calculations for the case involving both the electron-electron
and the electron-phonon interaction were based on classical, infinitely
heavy atoms and the condition $U \gg \lambda$. Here we want to point out that
there is another tractable
case, that is when the fluctuations for the combined system  are treated in a 
Gaussian approximation. Performing a Hubbard-Stratonovich transformation 
in Eq. (4) we obtain

\begin{eqnarray}
S_{eff,i}&=&\int^{\beta}_{0}d\tau \int^{\beta}_{0}d\tau'\sum_{\alpha=\uparrow
\downarrow}c_{i\alpha}^{*}(\tau)G_{0i\alpha}^{-1}(\tau-\tau')c_{i\alpha}(\tau')\nonumber\\
&-&\frac{1}{2} \int^{\beta}_{0}d\tau\,\phi_i(\tau)\Big(-\frac{d^{2}}{d\tau^{2}}+\omega_{0}^{2}\Big)\phi_i(\tau)\ \nonumber\\
&-&\int^{\beta}_{0}d\tau\, [\sqrt{U}y_i(\tau)-g\phi_i(\tau)][n_{i\uparrow}(\tau)-n_{i\downarrow}(\tau) ]\nonumber\\
&-&\frac{1}{2}\int^{\beta}_{0}d\tau y_i^2(\tau).
\end{eqnarray}
Changing variables as 
\begin{eqnarray}
\sqrt{U}x_i(\tau)=\sqrt{U}y_i(\tau)-g\phi_i(\tau),
\end{eqnarray}
and performing the Gaussian integral over the phonon field we obtain
 \cite{Nagaosa}
\begin{eqnarray}
S_{eff,i}&=&\int^{\beta}_{0}d\tau \int^{\beta}_{0}d\tau'\sum_{\alpha=\uparrow
\downarrow}c_{i\alpha}^{*}(\tau)G_{0i\alpha}^{-1}(\tau-\tau')c_{i\alpha}(\tau')\nonumber\\
&-&\int^{\beta}_{0}d\tau\, \sqrt{U}x_i(\tau)[n_{i\uparrow}(\tau)-n_{i\downarrow}(\tau) ]\nonumber\\
&-&\frac{1}{2}\sum_{n}\Big(\frac{\omega_n^2+\omega_0^2}{\omega_n^2+\Omega_0^2}\Big)x_{i}(-\omega_n)x_{i}(\omega_n),
\end{eqnarray}
where $\Omega_0=\omega_0\sqrt{1+\lambda /U}$ is the renormalized phonon 
frequency.
We now integrate out the electronic degrees of freedom. The effective action 
depends only on $x_{i}$ and it reads
\begin{eqnarray}
S_{x,i}&=&-\frac{1}{2}\sum_{n}\Big(\frac{\omega_n^2+\omega_0^2}
{\omega_n^2+\Omega_0^2}\Big)x_{i}(-\omega_n)x_{i}(\omega_n)\nonumber\\
&+& Tr\log[G_{0_{S \uparrow} }^{-1}-\sqrt{U}\,x_i]+Tr\log[G_{0_{S \downarrow}}^{-1}+\sqrt{U}\,x_i].
\end{eqnarray}
Expanding up to the first order in $\omega_n^2$ we obtain 

\begin{eqnarray}
S_{x,i}&=&-\frac{1}{2}\int^{\beta}_{0}d\tau \, 
x_i(\tau)\Big[ -\frac{1}{\omega_0^2(1+U/\lambda )}\frac{d^{2}}{d\tau^{2}}+1\Big]x_i(\tau)\nonumber\\
&+& Tr\log[G_{0_{S \uparrow} }^{-1}-\sqrt{U +\lambda}\,x_i] \nonumber\\
&+&Tr\log[G_{0_{S \downarrow}}^{-1}+\sqrt{U +\lambda}\,x_i].
\end{eqnarray}
The action Eq. (19) is no longer a function  of only the sum $U+\lambda$
but depends on both variables $U$ and $\lambda$ in an independent way. 

\begin{figure}
      \epsfysize=75mm
      \centerline{\epsffile{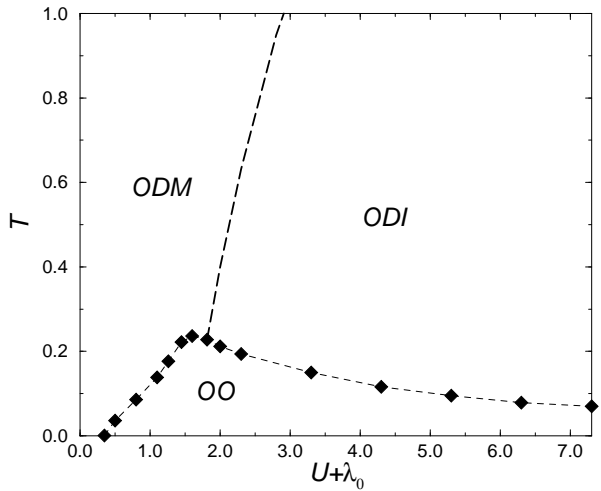}}
\narrowtext      
\caption{
Phase diagram showing the critical temperature   versus 
$U+ \lambda_0 $, with $\lambda_0=0.3$,  for dynamic phonons and HS fields  
in a Gaussian approximation.}
\label{fig:phase_b}
\end{figure}

The approximate action Eq. (19) reproduces correctly the cases without
phonons ($\lambda$ = 0) and without electronic correlations ($U$ = 0).
Moreover, for $T \gg \omega_0 \sqrt{1+U/\lambda}/2\pi$ only the zero-frequency
variable $x_i(0)$ contributes to the action corresponding
to the case where $x_i$ can be treated as a classical and time-independent
variable. Keeping all terms in Eq. (19) is equivalent to an Gaussian
approximation for temporal fluctuations. Eq. (19) therefore represents
an interpolation between the two exactly treated cases $\lambda = 0$ and 
$U = 0$ which is valid if the temperature is not much smaller than
the renormalized frequency $\omega_0 \sqrt{1+U/\lambda}/2\pi$. The
action given by Eq. (19) can be treated with the same methods as described
above for the phonon-only case. Figure \ref{fig:phase_b} shows the 
resulting phase
diagram for a fixed $\lambda =\lambda_0=0.3$  and $\omega_0 = 0.04$. 
 We have chosen $\lambda_0=0.3$ as a representative value for the 
electron-phonon coupling, according to what we have estimated above.
Comparison of
Figs. \ref{fig:phase_a} and  \ref{fig:phase_b} indicates that treating the atoms quantum-mechanically and
not classically with infinite heavy masses yields only small changes
justifying the above simpler treatment based on Eqs. (11)-(13).

\vspace{-0.5cm}

In conclusion, we have extended previous treatments of manganites
within the DMF  theory  including also phases
with long-range order and, in addition to the electron-phonon coupling,
the electron-electron interaction. Using expansions around
stationary paths of the action which are controlled by the Migdal parameter
we were able to treat the weak- and the strong-coupling cases on an equal
footing. However, since the temporal fluctuations
of the phonon and the Hubbard-Stratonovich (HS) fields are different
we considered three special cases: a) the phonon-only case;
b) the case where the phonons are treated statically, the HS field
dynamically; and c) treating both fields dynamically in the
Gaussian approximation. For each case the phase diagram was shown. 
Applying the results to undoped manganites it was concluded that the
observed Jahn-Teller transition is mainly driven by the 
electron-electron and only to a lesser extent by the electron-phonon
interaction. Our simple model for the Jahn-Teller transition in manganites
could be made more realistic by including additional interactions.
Anisotropic hopping terms, an electron-phonon interaction of second
order, and the explicit treatment of oxygen degrees of freedom would
remove the invariance of our Hamiltonian with respect to infinitesimal
rotations in orbital space and thus determine the electronic density
uniquely. Furthermore, instead of one effective electron-electron 
interaction constant, distinct inter- and intra- orbital interaction terms
 should be 
considered, 
which would allow to include also low-spin states.

{\bf Acknowledgment}: The authors  thank A. M. Ole\'s for a
careful reading of the manuscript and for discussions. 
P.B. acknowledges fruitful discussions with J. van den Brink.

\end{multicols}

\end{document}